\documentclass[aps,%
reprint,
superscriptaddress,
amsmath,amssymb,floatfix,
pre
]{revtex4-2}
\usepackage{lipsum}
\usepackage{hhline}
\usepackage{xtab,afterpage}
\usepackage{ upgreek }
\usepackage{mathrsfs}
\usepackage{booktabs}
\usepackage{pifont}
\usepackage{xcolor}

\usepackage{graphicx}
\usepackage{comment}
\usepackage{bm}
\usepackage{xcolor}
\usepackage{makecell}
\usepackage{amsmath,amsfonts,amssymb}
\DeclareMathAlphabet{\mathbbold}{U}{bbold}{m}{n}
\definecolor{darkblue}{rgb}{0,0,0.6}
\usepackage{hyperref}
\hypersetup{colorlinks,linkcolor=darkblue,citecolor=darkblue,urlcolor=darkblue}

\newcommand{\dd}{\text{d}}

\usepackage[normalem]{ulem}

\graphicspath{{figures/}}

\newcommand{\avg}[1]{\left\langle #1\right\rangle}

\newcommand{\Rc}{R_{\mathrm c}}

\newcommand{\kh}{\widehat{k}}

\newcommand{\Om}{\Omega_{d-1}}

\begin{document}
\title{Phase Separation in Fractonic Fluids: Coarsening, Interfaces, Nucleation, and Hyperuniformity in and out of Equilibrium}

\author{Raphaël Maire}
\email{maire@ub.edu}
\affiliation{Departament de Física de la Mat\`eria Condensada, Universitat de Barcelona, Martí Franqu\`es 1, 08028 Barcelona, Spain}

\begin{abstract}
   We develop a theory of phase separation in fluids conserving higher-order multipole moments of a scalar density, and investigate the resulting coarsening dynamics, nucleation, and interfacial fluctuations. We find that higher-order conservation laws slow the relaxation of all phenomena investigated, whereas nonequilibrium violations of the fluctuation--dissipation theorem suppress density and interfacial fluctuations while hindering nucleation. Simulations support these predictions.
\end{abstract}

\maketitle

\section{Introduction}
Fractonic phases are characterized by elementary excitations with restricted mobility, so that any transport requires collective rearrangements. First found in quantum models~\cite{chamon2005quantum,haah2011local,vijay2016fracton,yuan2020fractonic,stahl2022spontaneous,lake2022dipolar,stahl2023fracton}, fractonic behavior has since emerged as a broader paradigm for constrained many-body dynamics~\cite{nandkishore2019fractons,pretko2020fracton,gromov2024colloquium}, including the motion of topological defects in solids~\cite{pretko2018fracton, nguyen2020fracton}. These mobility constraints are naturally formulated through higher-rank gauge theories and symmetries associated to multipole conservation~\cite{pretko2017subdimensional,pretko2017generalized,gromov2019towards}.

Fractonic constraints can be imposed on either equilibrium or driven dynamics. For dynamics satisfying detailed balance, the stationary state follows the Gibbs distribution within each dynamically accessible sector~\cite{gliozzi2026domain}. Their approach to equilibrium may nonetheless be peculiar. Constraints can produce glassy relaxation~\cite{prem2017glassy} or prevent thermalization by splitting the space of configurations into disconnected sets~\cite{sala2020ergodicity,khemani2020localization,morningstar2020kinetically}. Where thermalization remains possible, hydrodynamic descriptions accounting for these higher-order conservation laws can describe the relaxation of long-wavelength perturbations, yielding, for instance, subdiffusion or other forms of slow dynamics~\cite{gromov2020fracton,feldmeier2020anomalous,iaconis2021multipole,iaconis2019anomalous,hart2022hidden,burchards2022coupled,mukherjee2024anomalous,mukherjee2023dynamic,hazra2025hyperuniformity} such as anomalous propagating modes when momentum is conserved~\cite{grosvenor2021hydrodynamics,glodkowski2023hydrodynamics,glorioso2023goldstone, glorioso2022breakdown,osborne2022infinite}. Such constrained transport can be realized in tilted lattices~\cite{zhang2020subdiffusion,guardadosanchez2020subdiffusion,scherg2021observing}.

Fractonic constraints also arise in classical nonequilibrium systems, where the dynamics can explicitly break time-reversal symmetry and the stationary state need not be Gibbsian. The hydrodynamics of such fluids is richer since relaxing this symmetry allows for additional relevant terms in the hydrodynamic equations~\cite{guo2022fracton}. Nonequilibrium classical particle or lattice systems that locally conserve both mass and center of mass, such as the biased random organization model~\cite{hexner2017noise,ma2019hyperuniformity}, the conserved Manna model~\cite{hexner2017noise}, the fixed-density Oslo model~\cite{mukherjee2024anomalous, hazra2025hyperuniformity} or interaction-mediated hopping~\cite{bertrand2019nonlinear} provide concrete examples of nonequilibrium fractonic fluids. As we will see below, such models and related ones~\cite{lei2019hydrodynamics,maire2025dynamical,lei2023random, lei2019nonequilibrium,kuroda2023microscopic,liu2025hyperuniform, li2025fluidization,kuroda2025singular, ikeda2024harmonic} display suppressed long-wavelength density fluctuations, a property called hyperuniformity, through the interplay between the conservation laws and nonequilibrium driving~\cite{lei2025non,maire2025hyperuniformity}. This suppression of fluctuations changes many collective phenomena relative to their equilibrium counterparts. The equilibrium conclusion of the Hohenberg--Mermin--Wagner theorem is evaded in such fractonic nonequilibrium crystals~\cite{galliano2023two,keta2025long,maire2024enhancing,kuroda2025long}, capillary waves and nucleation are suppressed~\cite{maire2025hyperuniform,maire2026hyperuniformity}, the liquid--gas critical point can lack critical opalescence and have an upper critical dimension reduced to $d_c=2$~\cite{gao2026liquid, ikeda2023correlated}, tracer dynamics is modified~\cite{alston2026stochastic} and entropy production is maximal~\cite{casiulis2026hyperuniform}. The effects of these constrained dynamics and fluctuation suppression have also been investigated in glasses~\cite{galliano2026glass, galliano2026identifying}. Taken together, these results suggest that center-of-mass conservation can qualitatively modify not only transport, but also the statistical mechanics of density fluctuations and phase-separated states. This motivates a coarse-grained field-theoretic treatment in which the conservation law can be systematically generalized to higher multipole moments.

Our goal is to develop such a coarse-grained field theory that extends these results for center-of-mass-conserving systems to generic fractonic fluids conserving higher mass moments. We focus primarily on phase separation and ask how these increasingly strong conservation laws modify its central statistical-mechanical phenomena: coarsening following spinodal decomposition, nucleation from a metastable phase, and fluctuations of the resulting interfaces. In doing so, we generalize the recent work of Ref.~\onlinecite{gliozzi2026domain}, which studied coarsening in fractonic fluids. To address these questions within a common framework, we work at the level of a coarse-grained hydrodynamic field theory for a scalar density $\rho$:
\begin{equation}
\partial_t\rho = -\Gamma (-\bm{\nabla}^2)^m \dfrac{\delta F}{\delta \rho}  +  \eta_n,
\label{eq:initial}
\end{equation}
where $F$ is an effective Landau free energy that allows phase separation. The power $m$ of the Laplacian and the spatial form of the noise enforce conservation of higher multipole moments. At equilibrium, this noise is constrained to satisfy the fluctuation--dissipation theorem (FDT) with respect to the mobility $\Gamma(-\bm{\nabla}^2)^m$. Far from equilibrium, however, this relation need not hold. Throughout, we therefore emphasize the difference between equilibrium and nonequilibrium fractonic systems.

In Sec.~\ref{sec:model}, we formulate the multipole-conserving stochastic field theory for an equilibrium and nonequilibrium fractonic fluid. In Sec.~\ref{sec:homogeneous}, we familiarize ourselves with Eq.~\eqref{eq:initial} by looking at some properties of homogeneous fractonic systems. In Sec.~\ref{sec:coarsening}, we allow for phase separation and analyze the coarsening dynamics and self-similar scaling of fractonic phase separating systems. In Sec.~\ref{sec:nucleation}, we investigate the nucleation processes in such systems while in Sec.~\ref{sec:interface}, we study the properties of interfaces between two phases. We conclude in Sec.~\ref{sec:conclusion}.

\section{Generalized stochastic gradient flow}
\label{sec:model}

\subsection{Multipole-conserving dynamics}

We denote by $\rho(\bm r,t)$ a conserved scalar density, which may be understood as a density deviation from a reference value and therefore need not be positive. For simplicity, we will call it the mass density, having in mind a fluid. Its $p$-th multipole moment is the symmetric rank-$p$ tensor
\begin{equation}
\mathcal Q^{(p)}_{i_1\cdots i_p}(t) = \int d \bm r r_{i_1}\cdots r_{i_p}\rho(\bm r,t).
\end{equation}
We require conservation of every component through order $P$:
\begin{equation}
\frac{d\mathcal Q^{(p)}_{i_1\cdots i_p}}{dt}=0, \qquad 0\leq p\leq P.
\end{equation}

Local dynamics implementing these constraints can be written in terms of a symmetric rank-$(P + 1)$ current~\cite{maire2025hyperuniformity,hexner2017noise,gliozzi2026domain}:
\begin{equation}
\partial_t\rho = \partial_{i_1}\cdots\partial_{i_{P + 1}} \mathcal J_{i_1\cdots i_{P + 1}},
\end{equation}
where repeated Cartesian indices are summed. Conservation follows by integration by parts, provided all boundary terms vanish.

We decompose the current into deterministic and fluctuating parts, $\mathcal J=\mathcal J^{\rm det} + \mathcal J^{\rm fluc}$. The simplest choice for the fluctuating current is additive Gaussian white noise, $\mathcal J^{\rm fluc}=\sqrt{2D} \xi$, where $\xi$ is a symmetric rank-$(P + 1)$ tensor with zero mean and covariance
\begin{equation}
\begin{split} 
  &\left\langle\xi_{i_1\cdots i_{P + 1}}(\bm r,t) \xi_{j_1\cdots j_{P + 1}}(\bm r',t')\right\rangle\\
&\qquad\qquad=\delta_{i_1(j_1}\cdots\delta_{i_{P + 1}j_{P + 1})} \delta(\bm r-\bm r')\delta(t-t'), 
\end{split}
\end{equation}
with parentheses denoting normalized symmetrization.

For the deterministic current, we take the simplest gradient expansion compatible with isotropy, including reflection symmetry. With only a scalar density, an even-rank current can be constructed from paired Kronecker deltas multiplying a function of $\rho$ while an odd-rank current requires one additional gradient. Each contracted pair of derivatives gives $\delta_{ij}\partial_i\partial_j=\bm{\nabla}^2$. With the additive current noise specified above, the most general equation compatible with these symmetries and conservation laws reads, to leading order in the gradient expansion,
\begin{equation}
\begin{split} \partial_t\rho={}& \begin{cases} (\bm{\nabla}^2)^{(P + 1)/2}A(\rho) + \mathcal O(\bm\nabla^{P + 3}\rho), & P\text{ odd},\\
(\bm{\nabla}^2)^{(P + 2)/2}B(\rho) + \mathcal O(\bm\nabla^{P + 4}\rho), & P\text{ even}, \end{cases}\\
& + \sqrt{2D} \partial_{i_1}\cdots\partial_{i_{P + 1}} \xi_{i_1\cdots i_{P + 1}},
\label{eq:isotropic_multipole_drift} \end{split}
\end{equation}
where $A$ and $B$ are arbitrary scalar functions. Here $\mathcal O(\bm\nabla^s\rho)$ denotes terms with at least $s$ spatial derivatives in total, including nonlinear combinations of density gradients.

At equilibrium, there is an additional constraint: the fluctuation--dissipation theorem relates the deterministic relaxation to the noise. For the same fluctuating current, it requires
\begin{equation}
\begin{split} \partial_t\rho={}&-\Gamma(-\bm{\nabla}^2)^{P + 1}C(\rho) + \mathcal O(\bm\nabla^{2P + 4}\rho)\\
& + \sqrt{2D} \partial_{i_1}\cdots\partial_{i_{P + 1}} \xi_{i_1\cdots i_{P + 1}},
 \end{split}
\end{equation}
where $C(\rho)$ is fixed by the equilibrium thermodynamics and $D=\Gamma T$, with Boltzmann's constant set to unity. Thus FDT requires $P + 1$ powers of the Laplacian in the deterministic term, rather than the lower powers allowed in Eq.~\eqref{eq:isotropic_multipole_drift}.

\subsection{Field equation and free energy}

We want to study phase separation, so we introduce the free energy
\begin{equation}
F[\rho]=\int\dd \bm r\left[f(\rho) + \frac{\kappa}{2}|\nabla\rho|^2\right],
\label{eq:F}
\end{equation}
where $f$ is a double-well potential and $\kappa>0$. To focus on the effects of noise and conservation, we retain deterministic dynamics generated by $F$. Nonvariational dynamics can produce interesting effects in their own right~\cite{solon2018generalized,omar2023mechanical,burekovic2026active}, such as microphase separation~\cite{tjhung2018cluster}. We omit these terms to isolate the effects of conservation and of the mismatch between deterministic transport and noise.

We therefore consider
\begin{equation}
\partial_t\rho=-\Gamma(-\bm{\nabla}^2)^m\frac{\delta F}{\delta\rho} + \eta_n,
\label{eq:model}
\end{equation}
with positive integers $m,n$, $\Gamma>0$, and zero-mean Gaussian noise
\begin{equation}
\avg{\eta_n(\bm r,t)\eta_n(\bm r',t')} =2D(-\bm{\nabla}^2)^n\delta(\bm r-\bm r')\delta(t-t'),
\end{equation}
or, equivalently, in Fourier space:
\begin{equation}
\avg{\eta_n(\bm k,t)\eta_n(\bm k',t')} =2Dk^{2n}(2\pi)^d\delta(\bm k + \bm k')\delta(t-t').
\end{equation}

The noise conserves every multipole through order $n-1$ while the deterministic term conserves every multipole through order $2m-1$, hence the full dynamics conserves every multipole through order
\begin{equation}
P=\min(2m-1,n-1),
\label{eq:Pmn}
\end{equation}
provided the corresponding boundary terms vanish. Our theory does not capture phase-space fragmentation. A density-dependent mobility with additional zero modes is one possible extension, motivated by microscopic fracton models~\cite{han2024scaling}.

For a highest conserved multipole order $P$, we choose the leading noise allowed by conservation, with exponent
\begin{equation}
n=P + 1.
\end{equation}
We will study two particular values of $m$:
\begin{equation}
m= \begin{cases} P + 1, & \text{at equilibrium},\\
\left\lceil (P + 1)/2 \right\rceil, & \text{out of equilibrium}. \end{cases}
\label{eq:m_as_a_function_of_p}
\end{equation}

At equilibrium, as we saw in the previous section, the fluctuation--dissipation relation requires~\cite{hohenberg1977theory}
\begin{equation}
m=n,\qquad D=\Gamma T.
\end{equation}

Out of equilibrium, we instead choose the minimal deterministic scaling compatible with isotropy and with conservation of multipoles through at least the same order $P$ as the noise which implies from Eq.~\eqref{eq:Pmn}:
\begin{equation}
m=\left\lceil \frac{n}{2}\right\rceil,
\end{equation}
where $\lceil a\rceil$ denotes the smallest integer greater than or equal to $a$. The ceiling is required in particular for even $P$, for which $n=P + 1$ is odd, since reflection symmetry forbids odd spatial derivatives of $\rho$.

For example, conservation of mass and center of mass corresponds to $P=1$, hence $n=2$. Writing the noise as $\sqrt{2D} \bm{\nabla}^2\xi$, where $\xi$ is a scalar Gaussian white noise with zero mean and unit spatiotemporal covariance, the equilibrium dynamics ($m=2$, $D=\Gamma T$) is
\begin{equation}
\partial_t\rho=-\Gamma(\bm{\nabla}^2)^2\frac{\delta F}{\delta\rho} + \sqrt{2\Gamma T}\bm{\nabla}^2\xi, \quad \text{(Equilibrium)}.
\end{equation}
Out of equilibrium, the leading deterministic term has $m=1$:
\begin{equation}
\partial_t\rho=\Gamma\bm{\nabla}^2\frac{\delta F}{\delta\rho} + \sqrt{2D}\bm{\nabla}^2\xi\quad \text{(Nonequilibrium)}.
\end{equation}
Both equations conserve mass and center of mass when boundary terms vanish. Their noise has the same Laplacian form, the lowest order allowed by these conservation laws. FDT then requires a fourth-order deterministic term at equilibrium, whereas a second-order term is allowed out of equilibrium. Thus, in a stable homogeneous phase away from criticality, the equilibrium fluid is subdiffusive, while the nonequilibrium fluid can be diffusive~\cite{hexner2017noise,maire2025hyperuniformity}. Saying that ``FDT fixes the noise'' can obscure this point: at fixed conservation laws, we can specify the leading noise first and use FDT to constrain the deterministic term.

One may ask whether such nonequilibrium deterministic terms, though allowed by the conservation laws, are generically present in fractonic systems. A pragmatic answer is that, for $n/2=m=1$, they underlie virtually all recently identified hyperuniform systems that do not require fine tuning~\cite{maire2025hyperuniformity,lei2025non}, suggesting that they are relatively generic. Ref.~\onlinecite{guo2022fracton} does not include such a low-order term in its general description of nonequilibrium fractonic fluids, but its analysis assumes that the stationary measure admits a local gradient expansion. Local dynamics, however, does not imply a local stationary measure~\cite{deluca2026generic}. In particular, our nonequilibrium theory does not satisfy this assumption since, as we will show, it displays hyperuniformity and therefore non-local correlations~\footnote{Technically, our Gaussian structure factor $S(\bm k)\sim k^2$ is analytic, however, its inverse is not.}.

Before analyzing the system, we wish to emphasize that we will use two complementary comparisons, depending on the observable of interest. First, we can hold the largest conserved multipole order $P$ fixed. Keeping the leading noise allowed by conservation gives
\begin{equation}
n=P + 1,\qquad m_{\rm eq}=P + 1,\qquad m_{\rm neq}=\left\lceil\frac{P + 1}{2}\right\rceil,
\end{equation}
where $m_{\rm neq}$ retains the leading symmetry-allowed deterministic term. The noise has the same spatial form in both cases, while the deterministic relaxation changes. This comparison is useful to see the difference between equilibrium and nonequilibrium systems \emph{at fixed} multipole conservation.

Alternatively, we can hold the deterministic dynamics fixed and change only the noise. For a given exponent $m$, we compare
\begin{equation}
n_{\rm eq}=m,\qquad n_{\rm neq}=\begin{cases} 2m, & P_{\rm neq}=2m-1,\\
2m-1, & P_{\rm neq}=2m-2. \end{cases}
\end{equation}
This comparison isolates the effect of noise while keeping the deterministic relaxation rates unchanged. It is therefore useful for identifying which results are controlled by deterministic relaxation and which depend on fluctuations. At fixed $m$, the conservation laws now differ between an equilibrium and a nonequilibrium system: the full dynamics conserves multipoles through $P_{\rm eq}=m-1$ at equilibrium, and through $2m-1$ or $2m-2$ in the two nonequilibrium cases. 

\section{Homogeneous-phase structure factor}
\label{sec:homogeneous}

\begin{figure*}
\centering
\includegraphics[width=0.9\linewidth]{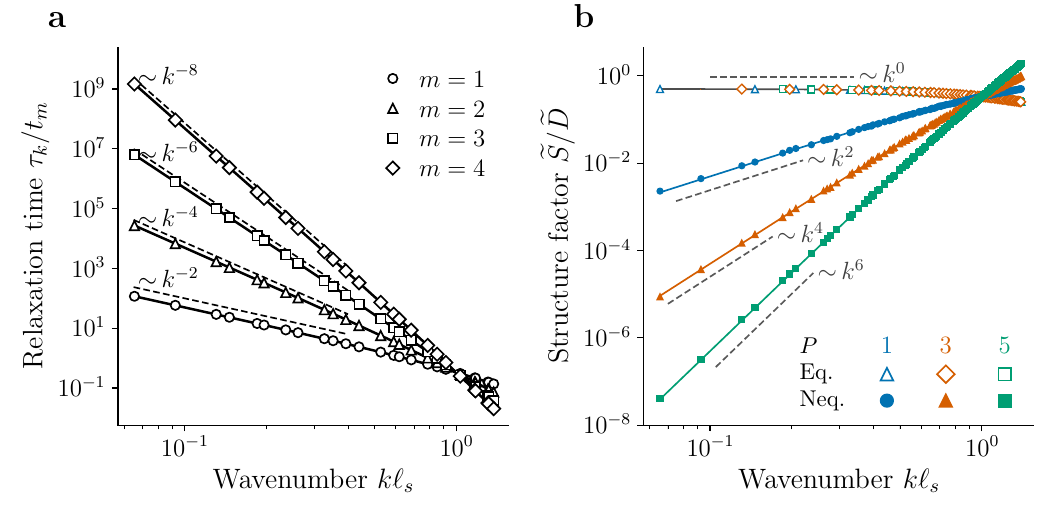}
\caption{Relaxation and stationary fluctuations about a homogeneous density $\rho_0=\rho_s$ in two dimensions. (a) Relaxation times of weak density perturbations under Eq.~\eqref{eq:model} with $D=0$. Solid curves show $\tau_k=1/\lambda_k$ from Eq.~\eqref{eq:lambda_k}. (b) Stationary structure factors at fixed conservation order. Solid curves follow Eq.~\eqref{eq:bulk_S}. All figures use $f(\rho)= f_s[(\rho/\rho_s)^2-1]^2/4$, where $\rho_s$ is a density scale and $ f_s$ a free-energy density scale. The length and time units are $\ell_s=\sqrt{\kappa\rho_s^2/ f_s}$ and $t_m=\rho_s^2\ell_s^{2m}/(\Gamma f_s)$. In $d=2$, $\widetilde S=S/(\rho_s^2\ell_s^2)$ and $\widetilde D=D/D_0$, with $D_0=\rho_s^2\ell_s^{2 + 2n}/t_m$. The grid spacing is $\ell_s$ and panel (b) uses $\widetilde D=10^{-3}$. With our choice of free energy, $\rho$ need not be positive and is therefore more naturally interpreted as a shifted density. Here we choose $\rho_0=\rho_s$. The equilibrium spectra are obtained from independent Gibbs samples.}
\label{fig:homogeneous}
\end{figure*}

We first investigate a system that does not phase separate and recover some results obtained in Ref.~\onlinecite{maire2025hyperuniformity}. We show that conservation of higher multipoles slows the relaxation of the density field and can suppress density fluctuations in nonequilibrium systems.

\subsection{Noncritical phase}
For small fluctuations about a homogeneous density $\rho_0$, let $\delta\rho_{\bm k}(t)=\int\dd\bm r e^{-\mathrm i\bm k\cdot\bm r}[\rho(\bm r,t)-\rho_0]$. We define the structure factor $S(\bm k)$ by $\langle\delta\rho_{\bm k}(t)\delta\rho_{\bm k'}(t)\rangle=(2\pi)^d\delta(\bm k + \bm k')S(k,t)$, or equivalently $S(k,t)=V^{-1}\langle|\delta\rho_{\bm k}(t)|^2\rangle$ in a periodic volume $V$. It quantifies the intensity of density fluctuations at a given scale $2\pi/|\bm k|$ and how fast they relax. From Eq.~\eqref{eq:model}, it is given by:
\begin{align}
S(k,t)={}&S_0(k)e^{-2\lambda_k t}  + \frac{D}{\Gamma}\frac{k^{2(n-m)}}{a + \kappa k^2} \left(1-e^{-2\lambda_k t}\right),\label{eq:bulk_time}\\
\lambda_k={}&\Gamma k^{2m}(a + \kappa k^2), \qquad a=f''(\rho_0)>0,\label{eq:lambda_k}
\end{align}
where $S_0(k)$ is the initial spectrum.

We verify our computation by weakly perturbing a homogeneous state and numerically solving the full nonlinear dynamics of Eq.~\eqref{eq:model} without noise using a spectral solver~\cite{caballero2024cupss}. Fig.~\ref{fig:homogeneous}(a) shows the relaxation times extracted from the decay of density-mode amplitudes. A density mode decays on the time scale $\tau_k=\lambda_k^{-1}$, which grows as $k^{-2m}$ at long wavelengths. Increasing $m$ therefore slows the relaxation of large-scale perturbations. This time scale depends on the deterministic dynamics and is independent of the noise exponent $n$. Using Eq.~\eqref{eq:m_as_a_function_of_p}, we find that at fixed highest conserved multipole order $P\geq1$, the leading nonequilibrium dynamics relaxes long-wavelength perturbations faster:
\begin{equation}
\tau_k^{\rm neq}\sim \begin{cases} k^{-(P + 1)}, & P\text{ odd},\\
k^{-(P + 2)}, & P\text{ even}, \end{cases} \qquad \tau_k^{\rm eq}\sim k^{-2(P + 1)}.
\end{equation}
The structure factor approaches its stationary value with the exponential $e^{-2\lambda_k t}$, hence on the time scale $\tau_k/2$. Conserving higher multipoles imposes stronger constraints and slows the dynamics, although relaxing the fluctuation--dissipation requirement allows faster relaxation at fixed $P$ since lower order deterministic terms are allowed.

Once a mode has relaxed and noise is included, its fluctuations reach the stationary spectrum shown in panel (b):
\begin{equation}
S(k,t\to\infty)=\frac{D}{\Gamma}\frac{k^{2(n-m)}}{a + \kappa k^2}.
\label{eq:bulk_S}
\end{equation}
At equilibrium, $n=m$ and $D=\Gamma T$, giving the Ornstein--Zernike form $S(k,t\to\infty)=T/(a + \kappa k^2)$. Its long-wavelength limit is finite and independent of $m$, as required by Gibbs statistics. Out of equilibrium, $m=\lceil n/2\rceil$ with $P\geq1$ instead gives a stationary state with suppressed long-wavelength density fluctuations ($S(k\to 0)\to 0$), a property called hyperuniformity~\cite{maire2025hyperuniformity}:
\begin{equation}
S(k,t\to\infty)\sim k^\alpha,\qquad \alpha=2(n-m)>0.
\end{equation}
For the nonequilibrium dynamics with $m=\lceil n/2\rceil$ at fixed $n=P + 1$,
\begin{equation}
\alpha=\begin{cases} P + 1=2m, & P\text{ odd},\\
P=2m-2, & P\text{ even}. \end{cases}
\label{eq:bulk_parity}
\end{equation}
Conserving higher multipoles therefore increases the suppression of large-scale fluctuations~\cite{maire2025hyperuniformity}. The mechanism leading to such suppression is the mismatch between the relaxation rate, proportional to $k^{2m}$, and the noise covariance, proportional to $k^{2n}$. At equilibrium, FDT requires $n=m$, so their ratio remains finite as $k\to0$. In the hyperuniform nonequilibrium case, $n>m$: noise vanishes faster than relaxation at long wavelengths, suppressing fluctuations as $S(k\to 0)\sim k^{2(n-m)}\to0$.

\subsection{Critical point}
At a mean-field liquid--gas critical point, the local free energy is tuned so that $f''(\rho_c)=f'''(\rho_c)=0$ and $f^{(4)}(\rho_c)>0$. In the Gaussian approximation, Eqs.~\eqref{eq:lambda_k} and \eqref{eq:bulk_S} then yield ($a=0$)
\begin{equation}
S_c(k)\sim k^{2(n-m-1)},\qquad \tau_{k,c}\sim k^{-(2m + 2)},
\label{eq:critical_scalings}
\end{equation}
corresponding to the Gaussian dynamic exponent $z_c^{(0)}=2m + 2$. Critical slowing down changes the relaxation-time scaling from $\tau_k\sim k^{-2m}$ away from criticality to $\tau_{k,c}\sim k^{-(2m+2)}$ at criticality.

Remarkably, unlike at an equilibrium critical point, density fluctuations need not diverge~\cite{gao2026liquid}. Indeed, for $n-m>1$, $S_c(k)$ instead vanishes as $k\to0$. Thus, sufficiently strong multipole conservation suppresses critical density fluctuations and hence critical opalescence. The static density response to a chemical-potential perturbation nevertheless diverges as $k^{-2}$~\cite{gao2026liquid}. Out of equilibrium, this is not inconsistent with a finite or vanishing structure factor because the fluctuation--dissipation theorem no longer relates the two. An effective relation can instead be restored by introducing a $k$-dependent temperature~\cite{gao2026liquid,maire2025dynamical,lei2019hydrodynamics}.

The Gaussian scaling in Eq.~\eqref{eq:critical_scalings} and our conclusions apply only above the upper critical dimension~\cite{gao2026liquid},
\begin{equation}
d_c=4-2(n-m).
\end{equation}
At equilibrium, where $n=m$, this reduces to the usual $d_c=4$, independently of the conservation laws. Out of equilibrium, with $m=\lceil n/2\rceil$, $d_c$ decreases with $n$ because the noise covariance, proportional to $k^{2n}$, becomes increasingly suppressed relative to the deterministic transport factor $k^{2m}$ at long wavelengths. For $n\geq4$, $d_c\leq0$, so Gaussian scaling holds in all physical dimensions. In particular, for the nonequilibrium $n=2$ case studied in Ref.~\onlinecite{gao2026liquid}, one finds $d_c=2$, with logarithmic corrections to scaling in $d=2$.

This upper critical dimension should not be confused with that of Ref.~\onlinecite{glorioso2022breakdown}, which arises from nonlinear hydrodynamic mode coupling in a fluid conserving both dipole moment and momentum, even away from criticality. The momentum mode driving that instability is absent from our scalar, overdamped theory.

\section{Self-similar phase separation}
\label{sec:coarsening}

We now choose a mean density and a free-energy density $f(\rho)$ that allow for phase separation through nucleation or spinodal decomposition. We focus first on the latter, where the homogeneous state $\rho(\bm r,t)=\rho_0$ is unstable and small perturbations grow whenever the relaxation rate in Eq.~\eqref{eq:lambda_k} becomes negative, i.e.,
\begin{equation}
f''(\rho_0) + \kappa k^2<0.
\end{equation}
For $f''(\rho_0)<0$, sufficiently long-wavelength modes therefore grow and initiate phase separation.

We show that conservation of higher multipoles slows coarsening and produces strong hyperuniformity in the self-similar regime, even for equilibrium dynamics. This hyperuniformity induced by large phase-separating patterns should not be conflated with the hyperuniformity in the homogeneous stationary state discussed above. The two are unrelated.

\begin{figure*}
\centering
\includegraphics[width=0.99\linewidth]{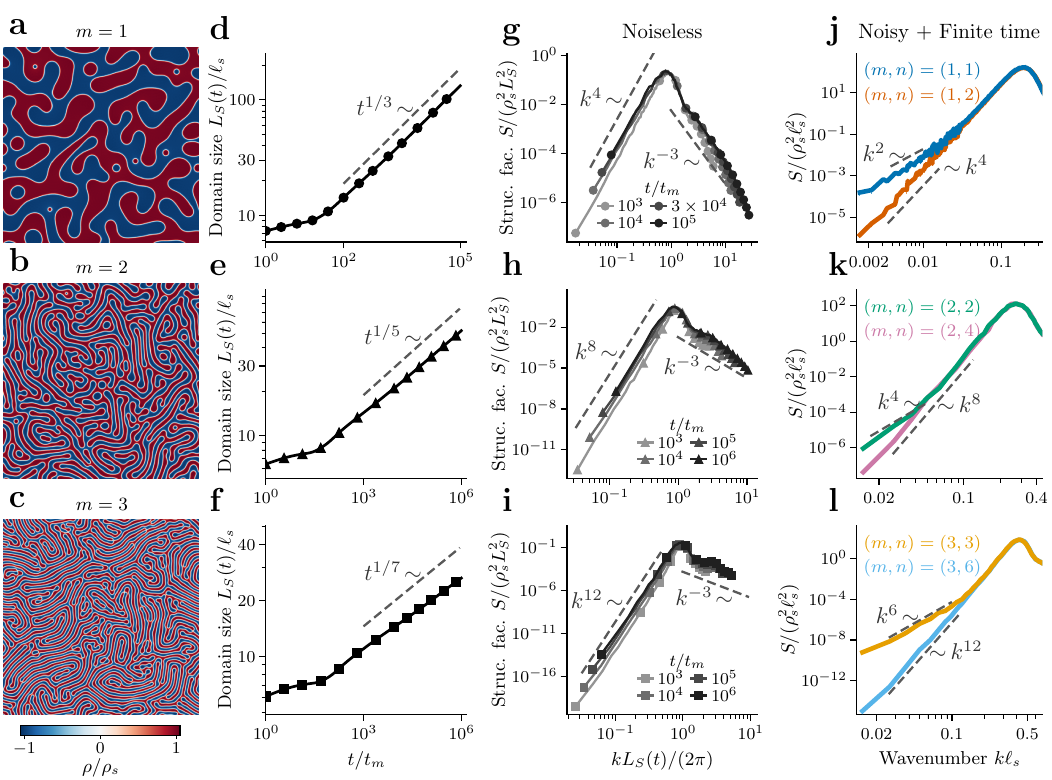}
\caption{Phase separation and coarsening in two dimensions under Eq.~\eqref{eq:model}, with $m=1,2,3$ from top to bottom. (a--c) Density fields $\rho(\bm r)/\rho_s$ at $t/t_m=10^4$, without noise. The mean density is zero. (d--f) Structure-factor domain size $L_S(t)/\ell_s$ as a function of time, starting from an almost homogeneous state. (g--i) Radially averaged spectra $S/(\rho_s^2L_S^2)$ versus $kL_S/(2\pi)$ at the indicated reduced times $t/t_m$, where $L_S=2\pi\int S(k) dk/\int kS(k) dk$ ($2\pi/L_S$ corresponds roughly to the peak of $S(k)$) and both integrals use $k\ell_s<1.45$. Panels (a--i) use $D=0$. (j--l) Spectra $\widetilde S$ versus $k\ell_s$ at fixed $m$ for equilibrium noise ($n=m$, $P=m-1$) and nonequilibrium noise ($n=2m$, $P=2m-1$). Panel (j) uses $t/t_m=10^3$ and $\widetilde D=0.03$; panels (k,l) use $t/t_m=10^4$, $\widetilde D=10^{-3}$, and three runs. Error bars in (j) are standard errors across runs. Grids contain $512^2$ points, except in (d,g), with $2048^2$, and (j), with $4096^2$. The free energy and units are those of Fig.~\ref{fig:homogeneous}.}
\label{fig:coarsening}
\end{figure*}

\subsection{Asymptotic growth law}

We start from a homogeneous system and let it evolve toward coexistence between two phases. Direct simulations of Eq.~\eqref{eq:model} without noise,
\begin{equation}
\partial_t\rho = -\Gamma(-\bm{\nabla}^2)^m \dfrac{\delta F}{\delta\rho},
\end{equation}
yield different patterns at finite $t$ as $m$ varies, as shown in Fig.~\ref{fig:coarsening}(a--c). Although the final state is fully phase separated, the transient dynamics depends on $m$.

Fig.~\ref{fig:coarsening}(d--f) shows that the characteristic domain size $L_S$ grows as a power law in time. Eventual saturation at the system size is not shown. For $m=1$, we recover the familiar diffusive coarsening law $L_S(t)\sim t^{1/3}$. This scaling applies both to equilibrium mass-conserving systems, such as Cahn--Hilliard dynamics~\cite{cahn1958free}, and to nonequilibrium fractonic systems conserving the center of mass: both have the same deterministic dynamics with $m=1$, and noise does not change the asymptotic growth exponent in this regime~\cite{bray1994theory}. Increasing $m$ slows coarsening, as also found in Ref.~\onlinecite{gliozzi2026domain}. Following Ref.~\onlinecite{bray1994theory}, we recover this result here for completeness.

If $L(t)$ denotes the characteristic domain size, the excess free energy comes from density gradients at the interfaces. The resulting chemical-potential scale is $\delta F/\delta\rho\sim \gamma/(\Delta\rho L)$, where $\Delta\rho$ is the coexistence density jump, $\gamma$ is the surface tension, which depends on $\kappa$ in Eq.~\eqref{eq:F} and $1/L$ is a curvature scale. The mass current $\bm J_m$, defined by $\partial_t\rho=-\bm \nabla\cdot \bm J_m$, satisfies
\begin{equation}
\bm J_m=-\Gamma\bm\nabla(-\bm{\nabla}^2)^{m-1}\dfrac{\delta F}{\delta\rho}\sim \frac{\Gamma\gamma}{\Delta\rho}L^{-2m}.
\end{equation}
Since domain growth proceeds through mass transport,
\begin{equation}
\dot L\sim \frac{\bm J_m\cdot\bm n}{\Delta\rho}, \qquad L(t)\sim\left[\frac{\Gamma\gamma t}{(\Delta\rho)^2}\right]^{1/(2m + 1)},
\end{equation}
in agreement with our numerical simulations and Ref.~\onlinecite{gliozzi2026domain}. In terms of the highest conserved multipole order $P$, using Eq.~\eqref{eq:m_as_a_function_of_p} yields:
\begin{equation}
L(t)\sim \begin{cases} t^{1/(2P + 3)}, & \text{equilibrium},\\
t^{1/(P + 2)}, & \text{nonequilibrium, odd }P,\\
t^{1/(P + 3)}, & \text{nonequilibrium, even }P. \end{cases}
\end{equation}
These laws assume a nondegenerate kinetic coefficient, no hydrodynamics, and a single growing scale. Local lattice dynamics with strict multipole constraints can display long preasymptotic regimes before the continuum exponent is visible~\cite{gliozzi2026domain}.

\subsection{Self-similar structure factor}

We now consider the structure factor of the evolving inhomogeneous fluid.

Motivated by equilibrium coarsening theory~\cite{bray1994theory} and the power-law growth derived above, we assume that the evolving density field becomes statistically self-similar at late times. Once initial transients have decayed, its spatial statistics should therefore become independent of time when lengths are rescaled by the characteristic domain size $L(t)$. This dynamic scaling hypothesis gives~\cite{bray1994theory}
\begin{equation}
S(k,t)=L(t)^d s(\kh),\qquad \kh=kL(t).
\end{equation}
The spectra collapse under this rescaling in Fig.~\ref{fig:coarsening}(g--i). The structure factor peaks at $k\sim 2\pi/L_S$ with an amplitude proportional to $L^d$, reflecting the characteristic domain size while sharp interfaces give the usual Porod tail at high $k$~\cite{bray1994theory}:
\begin{equation}
s(\kh)\sim \kh^{-(d + 1)}, \qquad 2\pi /L\ll k\ll 2\pi/\xi,
\end{equation}
independently of $m$ and with $\xi$ the interface width between domains. The small-$k$ behavior, however, is affected by $m$ and the system shows stronger and stronger hyperuniform scaling as $m$ is increased. We insist that this hyperuniformity is not related to the previously found hyperuniformity in the homogeneous regime. Rather, it is induced by the patterns selected by phase separation.

To understand the small-$\bm k$ behavior, we extend the deterministic argument of Refs.~\onlinecite{deluca2024hyperuniformity,tomita1991preservation} to general $m$. We first consider an initially almost perfectly homogeneous system $S(k\neq 0,0)\simeq 0$. We then ask what is the lowest order term in powers of $k$ that coarsening can induce. In Fourier space, each factor of $-\bm{\nabla}^2$ becomes $k^2$. Integrating the deterministic equation in time therefore gives
\begin{equation}
\rho_{\bm k}(t)-\rho_{\bm k}(0) =-\Gamma k^{2m}\int_0^t dt'\frac{\delta F}{\delta\rho(-\bm k, t')}.
\end{equation}
Thus every density change carries a factor $k^{2m}$. Since the structure factor measures the mean squared Fourier amplitude, the corresponding contribution carries a factor $k^{4m}$. No lower-order contribution can be generated by the dynamics and therefore:
\begin{equation}
s(\kh)\simeq A_{4m}\kh^{4m} + \dots, \qquad \kh=kL(t)\ll1,
\end{equation}
which is the scaling observed in Fig.~\ref{fig:coarsening}(g--i). For a more general initial spectrum, an initial background will remain at the smallest physical wave numbers, although it will generically be negligible in the late-time scaling limit of a large system~\cite{deluca2024hyperuniformity}.

Using Eq.~\eqref{eq:m_as_a_function_of_p}, the leading contribution can be written in terms of the highest conserved multipole order $P$ as
\begin{equation}
s(\kh)\sim \begin{cases} \kh^{4(P + 1)}, & \text{equilibrium},\\
\kh^{2(P + 1)}, & \text{nonequilibrium, odd }P,\\
\kh^{2(P + 2)}, & \text{nonequilibrium, even }P. \end{cases}
\end{equation}

\subsection{Noise-induced crossover in the structure factor}

So far, we have considered noiseless dynamics. In the coarsening regime studied here, weak noise leaves the growth law $L(t)\sim t^{1/(2m + 1)}$ unchanged, but can alter the small-$k$ structure factor. The reason is that noise injects fluctuations with a spectrum proportional to $k^{2n}$, whereas the deterministic dynamics contributes $k^{4m}$. At equilibrium, since $n=m$ and therefore $n<2m$, the noise contribution decays more slowly as $k\to0$ and can dominate at sufficiently small wave numbers, as illustrated in Fig.~\ref{fig:coarsening}(j--l). Following Ref.~\onlinecite{deluca2024hyperuniformity}, we estimate the time and scale at which a crossover between the two scalings is observed.

First, we identify the modes for which noise has not yet been balanced by relaxation. Using the bulk relaxation rate from Eq.~\eqref{eq:lambda_k}, with $a>0$ evaluated in a stable coexisting phase,
\begin{equation}
\lambda_{k_t}t=1, \qquad k_t\simeq(\Gamma a t)^{-1/(2m)}.
\end{equation}
The last expression applies at long wavelengths, where $\kappa k_t^2\ll a$. Modes with $k\ll k_t$ relax on time scales longer than $t$, so their noise-induced variance accumulates approximately linearly in time. The stochastic term in Eq.~\eqref{eq:bulk_time} then gives
\begin{equation}
S_{\rm inj}(k,t)\simeq \frac{Dk^{2n}}{\lambda_k}\left(1-e^{-2\lambda_k t}\right) \simeq 2Dtk^{2n},\qquad k\ll k_t.
\end{equation}
Adding this contribution to the rescaled structure factor gives
\begin{equation}
s(\kh,t)\simeq A_{4m}\kh^{4m}  + B_nDtL^{-(d + 2n)}\kh^{2n},
\end{equation}
where $B_n$ is a time-independent amplitude. For $n<2m$, equating the two terms defines the crossover $\kh_*=k_*L$:
\begin{align}
\kh_*&\sim\left(\frac{Dt}{L^{d + 2n}}\right)^{1/(4m-2n)},
\end{align}
Noise dominates below $\kh_*$, while $\kh^{4m}$ dominates above it. This estimate requires $k_*\ll k_t$ and $\kh_*\ll1$.

At equilibrium, $n=m$, so
\begin{align}
s(\kh,t)&\simeq A_{4m}\kh^{4m}  + B_mDtL^{-(d + 2m)}\kh^{2m},\\
\kh_*&\sim D^{1/(2m)}t^{(1-d)/[2m(2m + 1)]}.
\end{align}
For $d>1$, the noise-dominated region shrinks toward $\kh=0$ as time increases. The $\kh^{4m}$ law is therefore recovered at any fixed nonzero $\kh$ at late times, even though a lower-power tail can persist at finite times. For $m=1$ and $d=2$, this gives $\kh_*\sim D^{1/2}t^{-1/6}$, as found in Ref.~\onlinecite{deluca2024hyperuniformity}.

For the nonequilibrium dynamics, Eq.~\eqref{eq:m_as_a_function_of_p} gives two cases. When $P$ is odd, $n=2m$, and noise contributes the same power as the morphology:
\begin{equation}
s(\kh,t)\simeq \left[A_{4m} + B_{2m}DtL^{-(d + 4m)}\right]\kh^{4m}.
\end{equation}
Only the amplitude changes. The noise correction decays as $t^{(1-d-2m)/(2m + 1)}$, so it becomes negligible. When $P$ is even, $n=2m-1$ with $m\geq2$, and noise does generate a lower power:
\begin{align}
s(\kh,t)&\simeq A_{4m}\kh^{4m}  + B_{2m-1}DtL^{-(d + 4m-2)}\kh^{4m-2},
\\
\kh_*&\sim D^{1/2}t^{(3-d-2m)/[2(2m + 1)]}.
\end{align}
Hence, for all the cases considered here in $d\geq2$, noise becomes subleading at fixed nonzero $\kh$ in the late-time scaling limit, even when it conserves fewer multipoles than the deterministic dynamics. In $d=1$, noise can instead remain relevant, and the structure factor can follow $s(\kh, t)\sim \kh^{2n}$ even at late times~\cite{deluca2024hyperuniformity}.

Additional subtleties arise when the mobility depends on density or when the dynamics includes nonvariational terms that cannot be written as $\delta F/\delta\rho$. These extensions are discussed for $m=1$ in Ref.~\onlinecite{deluca2024hyperuniformity}.

\section{Nucleation}
\label{sec:nucleation}

We now consider a homogeneous state $\rho(\bm r)=\rho_0$ that is linearly stable, $f''(\rho_0)>0$, but metastable with respect to phase separation.

Our main finding will be that nucleation is more costly in the nonequilibrium systems considered here than at equilibrium because large-scale fluctuations are suppressed. This generalizes the result of Ref.~\onlinecite{maire2026hyperuniformity}.

\subsection{Evolution equation for the radius}

\begin{figure*}[t]
\includegraphics[width=\textwidth]{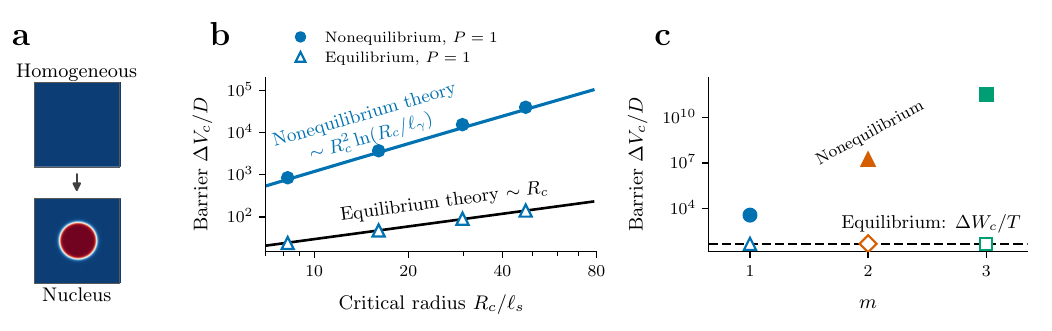}
\caption{Nucleation barriers in two dimensions. Barriers are plotted as $\Delta V_c/D$. Nonequilibrium symbols are finite-domain trial-action estimates obtained by the geometric minimum action method following Ref.~\onlinecite{zakine2024unveiling}. (a) A typical nucleation event from a homogeneous state. (b) Dependence of $\Delta V_c$ on $R_c=\gamma/\Delta\omega$ at fixed $P=1$, with $(m,n)=(1,2)$ out of equilibrium and $(2,2)$ at equilibrium. Increasing radii correspond to $\rho_0/\rho_s=-0.97,-0.985,-0.992,-0.995$ in reflecting disks with $R_{\mathrm{box}}/\ell_s=1536,1536,2048,8192$, respectively. The black solid curve is the equilibrium estimate $W(R_c)/T=\pi\gamma R_c/T$ while the blue solid curve is Eq.~\eqref{eq:nu_m1_barrier}. (c) Dependence on the nonequilibrium exponent $m$, with $n=2m$ and $P=2m-1$, for a fixed critical nucleus with measured finite-domain radius $R_c^{\rm box}/\ell_s=17.1903$ (the reservoir estimate is $R_c/\ell_s=16.0751$), at $\rho_0/\rho_s=-0.985$ in a disk with $R_{\mathrm{box}}/\ell_s=768$. The free energy and units are those of Fig.~\ref{fig:homogeneous}.}
\label{fig:nucleation}
\end{figure*}

We consider a homogeneous metastable dilute state of density $\rho_0$, weakly supersaturated relative to planar coexistence. Nucleation proceeds through a dense spherical nucleus of radius $R$ (see Fig.~\ref{fig:nucleation}(a)). We decompose the full conserved density as
\begin{equation}
\begin{gathered} \rho(\bm r,t)=g_{R(t)}(r) + \psi(\bm r,t),\\
g_R(r)\equiv g(r-R),\qquad r=|\bm r|. \end{gathered}
\end{equation}
Here $g$ is the mean-field planar coexistence profile derived in Appendix~\ref{app:nucleation_memory}. It smoothly interpolates between the two densities of the phase coexistence, accounting for the interface width in between the two. The remaining contribution to the full density field is denoted by $\psi$ which contains both supersaturation and all other density corrections.

Following Refs.~\onlinecite{cates2023classical,langford2025mechanics,chatzittofi2026nonequilibrium,ziethen2026nucleation,maire2026hyperuniformity}, we project the dynamics of the full system onto $R(t)$. Under the assumptions detailed in Appendix~\ref{app:nucleation_memory}, we obtain the dynamics of a nucleus of radius $R$ within the supersaturated medium:
\begin{equation}
\int_{-\infty}^t\dd t' Z_m(R_t,R_{t'};t-t')\dot R_{t'}=-W'(R_t) + \eta_{n,m}(R_t;t),
\label{eq:nucleation_radius}
\end{equation}
where $Z_m$ describes the memory of the conserved density field around a moving nucleus and $\eta_{n,m}(R_t;t)$ is a multiplicative Gaussian noise, with $R_t=R(t)$. The stochastic convention does not affect the leading weak-noise activation cost. At equilibrium, the fluctuation--dissipation theorem relates the noise correlation to $Z_m$.

The deterministic evolution of the nucleus radius is driven by the derivative of the reversible work of formation $W(R)$, which reflects the competition between the surface free-energy cost, set by the surface tension, and the bulk free-energy gain. For instance, in two dimensions,
\begin{equation}
W(R)=2\pi\gamma R-\pi\Delta\omega R^2,\qquad R_c=\frac{\gamma}{\Delta\omega},
\end{equation}
with surface tension $\gamma$ and bulk grand-potential gain
\begin{equation}
\Delta\omega=f(\rho_0)-f(\rho_{\rm d})-\mu_0(\rho_0-\rho_{\rm d})>0.
\end{equation}
Here $\rho_{\rm d}$ is the stable dense bulk at the same chemical potential as the metastable fluid: $f'(\rho_{\rm d})=f'(\rho_0)=\mu_0$. Due to the different scaling of the surface tension and bulk free-energy gain with $R$, the work of formation is nonmonotonic in $R$ with its maximum reached at the critical radius $R_c$. A droplet with radius below $R_c$ tends to shrink, while one above $R_c$ tends to grow. A rare fluctuation provided by the noise is therefore required for the formation of a droplet of radius $R>R_c$, before nucleation can proceed and the nucleus can grow deterministically.


\subsection{Activation cost and comparison with the figure}

In the weak-noise limit and starting from the metastable fluid, the probability of reaching the critical nucleus $R=R_c$ obeys the large-deviation form
\begin{equation}
P(R_c)\asymp \exp(-\Delta V_c/D).
\end{equation}
At equilibrium, $n=m=P + 1$ and $D=\Gamma T$. The noise and damping in Eq.~\eqref{eq:nucleation_radius} then satisfy the fluctuation--dissipation relation which guarantees that the (quasi-)stationary distribution is Gibbsian: $P(R)\sim \exp[-W(R)/T]$, so the large-deviation barrier is directly set by the reversible work:
\begin{equation}
\Delta V_c^{\rm eq}=\Gamma W(R_c).
\end{equation}
Conservation therefore modifies the dynamics and rate prefactor, but not the equilibrium nucleation barrier, provided the critical nucleus is dynamically accessible.

Out of equilibrium, $W$ no longer determines the activation cost by itself, and the metastable probability distribution must instead be obtained directly from Eq.~\eqref{eq:nucleation_radius}. Appendix~\ref{app:nucleation_memory} gives explicit barrier expressions in the regime $d>2m$ and a leading-logarithmic estimate for $d=2$, $(m,n)=(1,2)$. For the remaining nonlocal cases, large-$R_c$ scaling estimates can be obtained for $\Delta V_c$ but not exact formulas. In $d=2$ for the nonequilibrium cases we find:
\begin{align}
\Delta V_c &\simeq \frac{\pi\Gamma\kappa\Delta \rho^2}{3}\Rc^2\Lambda_c, &&P=1,\label{eq:nucleation_m1}\\[-2pt]
\Delta V_c &\propto\Rc^{(3P + 1)/2}, &&P\geq3\text{ odd},\\[-2pt]
\Delta V_c &\propto\frac{\Rc^{(3P + 2)/2}}{\mathcal L_c}, &&P\geq2\text{ even},
\end{align}
with $\Delta\rho$ the flat coexistence density difference. We used $n=P + 1$ and $m=\lceil n/2\rceil$. The factors $\Lambda_c$ and $\mathcal L_c$ grow as $\ln(R_c/\ell_\gamma)/2$ and are defined in Eqs.~\eqref{eq:nu_logarithm} and \eqref{eq:nu_even_logarithm}, with $\ell_\gamma=\chi\gamma/(\Delta\rho)^2$. Together with $\chi$ and $\Gamma$, this length controls the droplet response time and the spatial extent of mass redistribution~\footnote{A cruder projection that neglects mass redistribution would instead predict a response diverging with system size.}. For $P=1$, the memory admits a leading-logarithmic local approximation, giving an explicit estimate of $\Delta V_c$. For $P>1$, the stronger power-law memory must be retained, and we obtain only the scaling of $\Delta V_c$ with $R_c$. To test the theory, we numerically estimate $\Delta V_c$ by minimizing the Freidlin--Wentzell action of the full density field using the geometric minimum action method~\cite{zakine2023minimum,zakine2024unveiling,vandeneijnden2008geometric}. Fig.~\ref{fig:nucleation}(b) shows agreement with the equilibrium barrier and consistency with the nonequilibrium prediction for $(m,n)=(1,2)$ (Eq.~\eqref{eq:nucleation_m1}). At fixed $R_c$ and reversible work $W$, panel (c) shows an unchanged equilibrium barrier but an increasing nonequilibrium cost along $n=2m$. These fixed-radius comparisons do not directly test the predicted powers of $R_c$ for higher odd $P$ or even $P$. This trend indicates that conserving higher multipoles makes nucleation more costly. At large length scales, the nonequilibrium noise is weaker than the equilibrium noise, suppressing the density fluctuations needed to supply mass to a growing droplet. 
This comparison depends on the chosen noise normalization: although the barrier increases with $P$ at fixed $R_c$, the noise spectra scale differently with $k$, and no common effective temperature exists.

\section{Interfacial fluctuations}
\label{sec:interface}

\begin{figure*}[!ht]
\centering
\includegraphics[width=\textwidth]{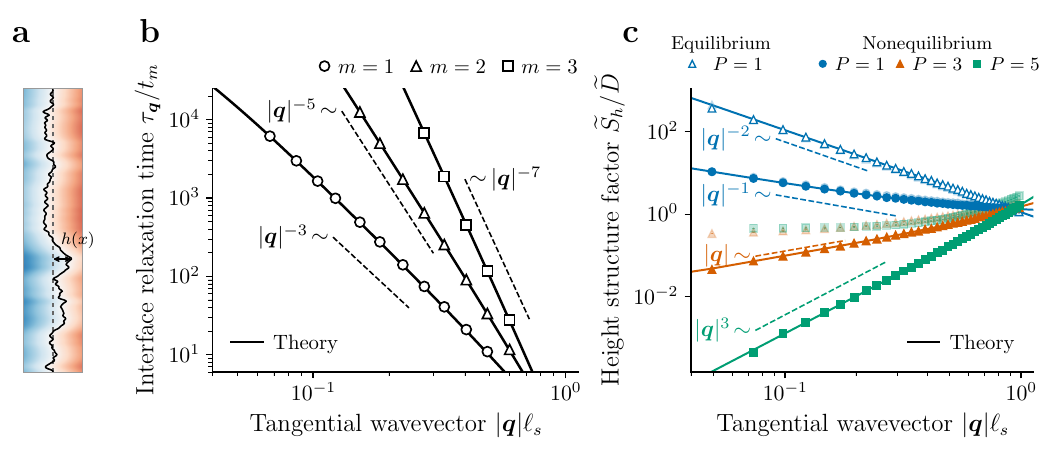}
\caption{Interfacial relaxation and fluctuations in two dimensions, using the free energy and units of Fig.~\ref{fig:homogeneous}. (a) Density snapshot with an interface. (b) Relaxation times $\tau_{\bm q}/t_m$ of weak interface perturbations versus $|\bm q|\ell_s$, for $m=1,2,3$ and $D=0$. Solid curves are the linear predictions for the simulated grid. (c) Profile spectra $\widetilde S_h/\widetilde D$ versus $|\bm q|\ell_s$, with $\widetilde S_h=S_h/\ell_s^3$. Opaque symbols show the spectrum when the interface height is determined by the same projection used theoretically to define $h$. Faint symbols use density-profile fits. The two methods for extracting $h$ are described in Appendix~\ref{app:howto}. Spectra use $256^2$ retained grids with spacing $\ell_s$, $L_x=L_y=256\ell_s$, and $\widetilde D=10^{-10}$. }
\label{fig:interface}
\end{figure*}

We have considered phase separation through spinodal decomposition and nucleation. We now investigate the final fate of these phenomena and look at the interface between two macroscopic phases, assuming that it is flat, as in a slab configuration, or that its radius of curvature is large enough for curvature effects to be negligible.

Our main findings will be that conservation laws slow the relaxation of interface modes, while nonequilibrium noise suppresses fluctuations around a flat interface. This generalizes the result of Ref.~\onlinecite{maire2025hyperuniform}.

\subsection{Projection onto a flat interface}

Consider a flat interface with normal coordinate $z$ and $d-1$ tangential coordinates $\bm x$ such that $\bm r=(z, \bm x)$ (see Fig.~\ref{fig:interface}(a)). We denote by $\bm q$ the $(d-1)$-dimensional wave vector conjugate to $\bm x$. The full bulk wave vector is $\bm k=(p,\bm q)$, where $p$ is conjugate to $z$. It is useful to decompose the density field as
\begin{equation}
\begin{gathered} \rho(\bm x,z,t)=g[z-h(\bm x,t)] + \psi(\bm r,t). \end{gathered}
\label{eq:interface_ansatz}
\end{equation}
As in the nucleation problem, $g(z-h)$ is the noiseless planar coexistence profile, centered at $z=h(\bm x,t)$, while $\psi$ contains the remaining density degrees of freedom. The projected displacement $h$ need not coincide with a local density contour when $\psi\neq0$.

Using standard methods~\cite{sarfati2026from,despeignes2026interface,besse2023interface,langford2024theory}, detailed in Appendix~\ref{app:interface_constants}, we obtain from Eq.~\eqref{eq:interface_ansatz} an evolution equation for the interface on large scales:
\begin{equation}
\begin{gathered} \partial_t h(\bm q, t)=-\nu ^{h}|\bm q|^{2m + 1}h(\bm q,t) +  \sqrt{2 \mathcal D_{m,n}^{h} |\bm q|^{\alpha_{m, n}}}\eta(\bm q, t).\\
\alpha_{m, n}= \begin{cases} 2n-1, & n\leq 2m-1,\\
4m-2, & n\geq 2m, \end{cases} \end{gathered}
\label{eq:interface_projected}
\end{equation}
Here $\eta$ is unit Gaussian white noise along the interface and in time. The coefficients $\nu ^{h}$ and $\mathcal D_{m,n}^{h}$, independent of $\bm q$, are given in Appendix~\ref{app:interface_constants}. Unlike the nucleation problem, the interface dynamics does not require a long-time memory kernel. Indeed, bulk modes relax as $|\bm q|^{2m}$, whereas interface modes relax more slowly, as $|\bm q|^{2m + 1}$. This separation of time scales allows a local-in-time equation for the capillary mode at leading order in the long-wavelength limit. Droplet growth, by contrast, couples through the mass redistribution to arbitrarily long-wavelength bulk modes, whose elimination generally generates memory. The underlying conservation laws nevertheless remain manifest in the spatially nonlocal dynamics of $h(\bm x)$. Indeed, the relaxation rate $|\bm q|^{2m + 1}$ is nonanalytic in $\bm q$ and therefore cannot be represented by a finite-order local differential operator in real space, unlike $|\bm q|^{2m}$, which corresponds to a local operator proportional to $(-\bm{\nabla}^2)^m$. This non-locality arises from the integration of the underlying conserved dynamics~\cite{onuki2002phase}.

The height structure factor is defined by $\langle h(\bm q,t)h(\bm q',t)\rangle=(2\pi)^{d-1}\delta(\bm q + \bm q')S_h(\bm q,t)$, with $S_{h,0}(\bm q)$ its initial value. Eq.~\eqref{eq:interface_projected} gives
\begin{equation}
\begin{aligned} S_h(\bm q,t)=&\frac{\mathcal D_{m,n}^{h}}{\nu ^{h}}|\bm q|^{\alpha_{m,n}-2m-1} \left(1-e^{-2 t/\tau_{\bm q}}\right)\\
 + &S_{h,0}(\bm q)e^{-2t/\tau_{\bm q}}, \quad \tau_{\bm q}=\frac{1}{\nu ^{h}|\bm q|^{2m + 1}}. \end{aligned}
\label{eq:Sh_time}
\end{equation}
The height-mode amplitude relaxes on the time scale $\tau_{\bm q}$, while its structure factor approaches stationarity on the time scale $\tau_{\bm q}/2$. The predicted mode relaxation time agrees well with the decay measured in noiseless phase-separated systems following a weak perturbation of the interface, as shown in Fig.~\ref{fig:interface}(b). Again, increasing the number of conservation laws leads to progressively longer relaxation times.

Once a mode has relaxed, Eq.~\eqref{eq:Sh_time} approaches
\begin{equation}
\begin{gathered} S_h(\bm q,t\to\infty)=\frac{\mathcal D_{m,n}^{h}}{\nu ^{h}}|\bm q|^{\beta_{m,n}},\\
\beta_{m,n}=\begin{cases} 2(n-m)-2,&n\leq2m-1,\\
2m-3,&n\geq2m. \end{cases} \end{gathered}
\end{equation}
These expressions apply at wavelengths large compared with the intrinsic interface width. At equilibrium, $n=m$ and $D=\Gamma T$, and the coefficients in Appendix~\ref{app:interface_constants} yield
\begin{equation}
S_h^{\rm eq}(\bm q,t\to\infty)=\frac{T}{\gamma |\bm q|^2},
\end{equation}
which is the standard capillary-wave result.

For the nonequilibrium dynamics with $m=\lceil n/2\rceil$, we obtain
\begin{equation}
\begin{gathered} S_h^{\rm neq}(\bm q,t\to\infty)\sim |\bm q|^{P-2}, \end{gathered}
\end{equation}
This is the interface counterpart of Eq.~\eqref{eq:bulk_parity}. The mismatch between noise and dissipation suppresses long-wavelength height fluctuations relative to the equilibrium $|\bm q|^{-2}$ spectrum~\cite{maire2025hyperuniform}. For $P=1$, the spectrum still diverges as $|\bm q|^{-1}$, for $P=2$, it tends to a constant and for $P\geq3$, it vanishes as $|\bm q|\to0$.

Fig.~\ref{fig:interface}(c) compares the predicted spectra with two measurements of the interface height. For the opaque symbols, $h$ is measured using the leading-order projected definition adopted in the theory (see Appendix~\ref{app:howto}), yielding agreement for all tested $P$. Alternatively, one may define $h$ by fitting the density near the interface to a translated planar profile with fixed width and amplitude. The two definitions yield essentially identical $S_h$ when the spectrum remains divergent as $q\to0$, but differ when $S_h$ is predicted to vanish at small $|\bm q|$. Indeed, we see that for the second method of measuring $h(\bm x)$, the spectrum for the fitted height instead develops a low-$q$ plateau. In this case, density fluctuations near the interface can shift the fitted position even without displacing the underlying profile, so that the fitted height also responds to the residual fluctuations $\psi$ in Eq.~\eqref{eq:interface_ansatz}. These residual fluctuations are eliminated to leading order in the construction of $h(\bm x)$ with the first method. Hence, this contribution can mask the suppression of $S_h$ at small $|\bm q|$. Thus, when higher-multipole conservation causes the low-$q$ power to vanish, the definition of the interface position becomes important.

\section{Conclusion}
\label{sec:conclusion}

We developed a field theory of phase separation with higher multipole conservation and separated the effects of constrained transport from those of a broken fluctuation--dissipation relation. Higher multipole conservation slows coarsening and interfacial relaxation. Out of equilibrium, the mismatch between noise and dissipation suppresses large-scale density and height fluctuations and increases the activation cost of nucleation.

Further work could explore crystalline order in nonequilibrium fractonic systems, where the suppression of long-wavelength phonons can stabilize long-range translational order beyond the equilibrium Hohenberg--Mermin--Wagner constraint~\cite{galliano2023two,maire2024enhancing,ikeda2024harmonic}. This mechanism is distinct from the translation-symmetry breaking induced by phase-space fragmentation and nonergodicity in Hamiltonian fracton models~\cite{sadki2026statistical}. Developing a theory of nonequilibrium melting in such systems, particularly in $d=2$, is also an interesting direction.

Our simplified field theory does not capture phase-space fragmentation. Its effect on the predicted scaling could be investigated using a mobility with additional zero modes or particle and lattice models that exhibit nonergodic fractonic behavior, for instance by extending the models of Refs.~\onlinecite{gliozzi2026domain, han2024scaling} to admit phase separation.

Another direction is to test the robustness of these results to nonvariational terms in the density dynamics, as studied in active matter~\cite{cates2025active}.

\section*{Acknowledgments}
I thank Andrew Lucas for helpful exchanges regarding time-reversal symmetry and nonlocal steady-state correlations in fractonic hydrodynamics.

\appendix

\section{Adjoint projection of the conserved nucleation dynamics}
\label{app:nucleation_memory}

We adapt the collective-coordinate construction of Ref.~\onlinecite{chatzittofi2026nonequilibrium} to define the droplet radius. In contrast to their nonconserved case, no localized adjoint weight can eliminate the residual field from the conserved dynamics. We therefore work at the level of the transport equation, drawing on Ref.~\onlinecite{elder2001sharp}.

\subsection{Density decomposition and definition of the radius}

We start from
\begin{equation}
\begin{gathered} \dot\rho=-\Gamma A^m\mu[\rho] + \eta_n,\\
\mu[\rho]=f'(\rho) + \kappa A\rho, \qquad A=-\bm{\nabla}^2, \end{gathered}
\label{eq:nu_starting_field}
\end{equation}
with zero-mean Gaussian noise of covariance
\begin{equation}
\langle\eta_n(\bm r,t)\eta_n(\bm r',t')\rangle = 2D A^n \delta(\bm r-\bm r')\delta(t-t').
\end{equation}

Assuming the existence of a nearly spherical nucleus of radius $R(t)$, we decompose
\begin{equation}
\rho(\bm r,t)=g_{R(t)}(r) + \psi(\bm r,t), \qquad g_R(r)=g(r-R),
\label{eq:nu_density_split}
\end{equation}
where $g$ is the planar coexistence profile,
\begin{equation}
f'(g)-\kappa g''=\mu_{\rm coex}, \qquad g(-\infty)=\rho_ + , \quad g( + \infty)=\rho_-.
\end{equation}

The function $g_R$ uses the planar mean-field profile centered at $r=R$. All corrections to it are contained in $\psi$, including curvature corrections, the supersaturated background, fluctuations, and the depletion or accumulation halo around the droplet itself, as required by mass conservation.

We define
\begin{equation}
\Delta\rho=\rho_ + -\rho_-, \qquad \mu_0=f'(\rho_0), \qquad \delta\mu=\mu_0-\mu_{\rm coex}.
\end{equation}
For weak supersaturation, the difference between the metastable homogeneous density and the dilute coexistence density imposed by $g$ is
\begin{equation}
\delta\rho_b\equiv\rho_0-\rho_- = \chi_-\delta\mu + \mathcal O(\delta\mu^2), \qquad \chi_\pm=\frac1{f''(\rho_\pm)}.
\label{eq:nu_supersaturation_offset}
\end{equation}
For simplicity we use a common bulk susceptibility
\begin{equation}
\chi=\chi_-=\chi_ + .
\end{equation}
Thus $\psi$ also contains this background contribution (since $g(\infty)=\rho_-$).

Differentiating Eq.~\eqref{eq:nu_density_split} gives
\begin{equation}
\partial_t\rho = -g'_R\dot R + \partial_t\psi, \qquad g'_R(r)=g'(r-R).
\label{eq:to_project}
\end{equation}
The decomposition $\rho=g_R + \psi$ remains arbitrary until the radius $R$ is defined. We fix $R$ by requiring the residual field $\psi$ to be orthogonal to an interfacial weight:
\begin{equation}
\langle q_R,\psi\rangle=0, \qquad \langle a,b\rangle\equiv\int d \bm r a^*b,
\end{equation}
where $q_R$ is localized near the interface, has nonzero overlap with the radius tangent $-g'_R$, and will be determined below.

Using Eq.~\eqref{eq:to_project} and the Stratonovich convention,
\begin{equation}
\dot R = \frac{\langle q_R,\partial_t\rho\rangle} {\langle q_R,-g'_R\rangle-\langle\partial_Rq_R,\psi\rangle}.
\label{eq:nu_radius_identity_general}
\end{equation}
The evolution equation also gives, to linear order in $\psi$,
\begin{equation}
\langle q_R,\partial_t\rho\rangle \simeq-\Gamma\langle A^m q_R,b_R\rangle -\Gamma\langle H_R A^m q_R,\psi\rangle  + \langle q_R,\eta_n\rangle,
\label{eq:222}
\end{equation}
with
\begin{equation}
\begin{gathered} b_R(r)\equiv\mu[g_R](r)-\mu_0=-(\mu_0 - \mu_{\rm coex})-\frac{(d-1)\kappa}{r}g'_R(r).\\
H_R=f''(g_R) + \kappa A. \end{gathered}
\end{equation}
At this point the derivation differs from its nonconserved counterpart. In Ref.~\onlinecite{chatzittofi2026nonequilibrium}, $q_R$ is chosen to satisfy $H_RA^mq_R=0$ to eliminate $\psi$ from Eq.~\eqref{eq:222}. For a nonconserved model ($m=0$), $H_R$ is self-adjoint and an approximate solution is $q\propto g'$ because $H_Rg'\simeq H_g g'=0$. The planar operator $H_g=f''(g)-\kappa\partial_z^2$ differs from the curved operator $H_R$ by a small curvature correction. Attempting the analogous cancellation for conserved dynamics ($m\geq1$) would formally require
\begin{equation}
A^m q_R\propto g'_R,
\end{equation}
to satisfy $H_RA^mq_R=0$. For conserved dynamics, however, this equation has no admissible localized solution. Indeed, a change in the reference droplet radius, $\delta\rho_{\rm core}=-g'_R\delta R$, changes its mass,
\begin{equation}
\delta M_{\rm core} =-\delta R\int d \bm r g'_R\neq0,
\label{eq:mass}
\end{equation}
whereas for periodic or no-flux boundary conditions every field in the range of $A^m$ has zero spatial integral,
\begin{equation}
\int d\bm r A^m q_R=(-1)^m\int d\bm r \bm{\nabla}^{2m} q_R=0,
\label{eq:divergence}
\end{equation}
by the divergence theorem. Hence, $A^mq_R\propto g'$ cannot work because Eq.~\eqref{eq:divergence} is not compatible with Eq.~\eqref{eq:mass}. Thus $g'_R$ is not in the range of $A^m$ and formally writing $q\propto A^{-m}g'_R$ either requires subtracting the conserved zero mode, in which case $A^m q$ is no longer proportional to $g'_R$, or, in an infinite reservoir, gives a long-ranged weight whose static normalization may diverge in the dimensions discussed below. Physically, this reflects the fact that a change of radius $R$ cannot occur without a compensating change of mass in $\psi$:

\begin{equation}
\delta\rho=-g'_R\delta R + \delta\psi, \qquad \int d \bm r\delta\psi = \delta R\int d\bm r g'_R,
\end{equation}

The compensating part belongs to $\psi$ and is not eliminated by this attempted static interface-mode projection. We must retain its dynamics, which generate memory in the effective equation for $R(t)$ and account for the redistribution of mass and higher multipole moments when the radius changes.

The evolution equation for the density field gives
\begin{equation}
(\partial_t + \Gamma A^mH_R)\psi = g'_R\dot R -\Gamma A^m b_R  + \eta_n  + \mathcal O(\psi^2).
\label{eq:nu_residual_linear}
\end{equation}
Together with Eq.~\eqref{eq:nu_radius_identity_general}, this specifies the linearized dynamics. Although Eq.~\eqref{eq:nu_residual_linear} can be solved formally, its dependence on the full history of $b_R$ complicates the calculation. We therefore take another route.

\subsection{Slow bulk chemical-potential response}

The exact density evolution, Eq.~\eqref{eq:nu_starting_field}, together with the decomposition in Eq.~\eqref{eq:nu_density_split}, can be rewritten as
\begin{equation}
\begin{split} \chi(\partial_t + \nu A^m)\mu = \left(g'_R\dot R + \eta_n\right)- \partial_t \left[ \psi-\chi(\mu-\mu_{\rm coex}) \right], \end{split}
\label{eq:nu_transport_exact}
\end{equation}
where $\nu=\Gamma/\chi$. We close this equation by assuming that, on the time scale of nucleus growth, the bulk density follows the local susceptibility relation in Eq.~\eqref{eq:nu_supersaturation_offset}:
\begin{equation}
\partial_t\left[\psi-\chi(\mu-\mu_{\rm coex})\right]\simeq0.
\end{equation}
Thus the slow part of $\partial_t\psi$ is approximated by $\chi\partial_t\mu$, while curvature corrections and other distortions of the interfacial profile are taken to relax quasistatically. This closure is appropriate in the thin-interface regime when the nucleus evolves slowly compared with these local relaxation processes. Eq.~\eqref{eq:nu_transport_exact} therefore reduces to
\begin{equation}
(\partial_t + \nu A^m)\mu \simeq \frac1\chi g'_R\dot R  + \frac1\chi\eta_n.
\label{eq:nu_mu_transport}
\end{equation}

For an initial time $t_0$ with $\mu(t_0)=\mu_0$, a formal solution to Eq.~\eqref{eq:nu_mu_transport} is
\begin{equation}
\mu(t) \simeq \mu_0 +  \frac1\chi \int_{t_0}^t dt' E_m(t-t')\left(g'_{R_{t'}}\dot R_{t'} + \eta_n(t')\right),
\label{eq:nu_mu_solution}
\end{equation}
with
\begin{equation}
E_m(\tau)=e^{-\nu A^m\tau}.
\end{equation}

We now project Eq.~\eqref{eq:nu_mu_solution} with the interfacial weight $q_{R_t}$ at the observation time $t$:
\begin{equation}
\begin{split} \langle q_{R_t},(\mu(t) - \mu_0)\rangle \simeq&- \int_{t_0}^t dt' Z_m(R_t,R_{t'};t-t')\dot R_{t'}\\
& +  \eta_{n,m}(R_t;t), \end{split}
\label{eq:nu_mu_projected_solution}
\end{equation}
where
\begin{equation}
\begin{gathered} Z_m(R,R';\tau) \equiv- \frac1\chi \langle q_R,E_m(\tau)g'_{R'} \rangle,\\
\eta_{n,m}(R_t;t) \equiv \frac1\chi \int_{t_0}^t dt' \langle q_{R_t}, E_m(t-t')\eta_n(t') \rangle. \end{gathered}
\end{equation}

Eq.~\eqref{eq:nu_mu_projected_solution} establishes a dynamical relation between the history of the radius and the chemical potential. The nonlocal kernel appears because a displacement of the droplet wall requires redistribution of a conserved density through the surrounding bulk. The projector $q_R$ is still arbitrary.

\subsection{Interfacial matching and the capillary force}

We now obtain an independent expression for the same projected chemical potential from the narrow interfacial region.

Expanding the chemical potential to first order in $\psi$ and curvature and collecting the terms involving $\psi$ gives
\begin{equation}
H_g\psi(R + z,t) \simeq \mu(R + z,t)-\mu_{\rm coex}  +  \frac{(d-1)\kappa}{R}g'(z),
\label{eq:nu_inner_expansion}
\end{equation}
where
\begin{equation}
H_g=f''(g)-\kappa\partial_z^2, \qquad z=r-R(t).
\end{equation}

Although we seek $\mu$, Eq.~\eqref{eq:nu_inner_expansion} must also admit a solution for $\psi$. We cannot directly invert $H_g$ because it has a zero mode, $H_g g'=0$. We therefore impose a Fredholm solvability condition on the right-hand side of Eq.~\eqref{eq:nu_inner_expansion}:
\begin{equation}
\int dz g'(z)\left(\mu(R + z,t)-\mu_{\rm coex}  +  \frac{(d-1)\kappa}{R}g'(z)\right) = 0.
\end{equation}
This gives an independent expression for the projected chemical potential:
\begin{equation}
\begin{split} \langle g'_R,(\mu - \mu_0)\rangle\!=\! \Omega_{d-1}R^{d-1} \left[ \Delta\rho(\mu_0-\mu_{\rm coex})-\frac{(d-1)\gamma}{R} \right]\!, \end{split}
\label{eq:nu_force_projection_pre}
\end{equation}
Here $\Omega_{d-1}$ is the surface area of the unit $(d-1)$-sphere and $\gamma=\kappa\int dz (g'(z))^2$ is the surface tension. Eq.~\eqref{eq:nu_force_projection_pre} is the Gibbs--Thomson relation. For weak supersaturation,
\begin{equation}
\Delta\omega = \Delta\rho(\mu_0-\mu_{\rm coex})  + \mathcal O(\delta\mu^2),
\end{equation}
so that
\begin{equation}
\langle-g'_R,(\mu-\mu_0)\rangle \simeq W'(R),
\label{eq:nu_force_projection}
\end{equation}
with $W(R)$ the reversible cost of formation:
\begin{equation}
W(R) = \Omega_{d-1} \left[ \gamma R^{d-1} -\frac{\Delta\omega}{d}R^d \right].
\end{equation}

\subsection{Effective radius dynamics}

Choosing $q_R=-g'_R$ gives two expressions for the same projected chemical-potential difference. Combining Eq.~\eqref{eq:nu_force_projection} and Eq.~\eqref{eq:nu_mu_projected_solution} yields
\begin{equation}
\int_{t_0}^t dt' Z_m(R_t,R_{t'};t-t')\dot R_{t'} = -W'(R_t)  +  \eta_{n,m}(R_t;t).
\end{equation}

We emphasize that $q_R$ is not fixed by a projection of the full dynamics, as in Ref.~\onlinecite{chatzittofi2026nonequilibrium}. Instead, it is determined by requiring that the projected chemical potential admit an independent evaluation, which in turn selects a specific projection profile $g'_R$.

\subsection{Memory kernel and projected noise}
The memory kernel
\begin{equation}
Z_m(R,R';\tau) =\frac{1}{\chi}\left\langle g'_R, e^{-\nu A^m\tau}g'_{R'}\right\rangle, \quad \nu =\frac{\Gamma}{\chi},
\label{eq:nu_memory_kernel_compact}
\end{equation}
and, in the stationary limit $t_0\to-\infty$, the noise-field covariance at prescribed radii $R,R'$, with $\tau=t-t'$,
\begin{equation}
\begin{split} & B_{m,n}(R,R';\tau)=\left\langle\eta_{n,m}(R;t)\eta_{n,m}(R';t')\right\rangle\\
&\qquad=\frac{D}{\Gamma\chi} \left\langle g'_R, A^{n-m}e^{-\nu A^m|\tau|}g'_{R'}\right\rangle, \end{split}
\label{eq:nu_noise_time}
\end{equation}
satisfy the fluctuation--dissipation theorem at equilibrium when $n=m$, with $B_{m,m}=T Z_m$ and $T=D/\Gamma$. The radius equation samples this noise field at $R=R_t$.

\subsection{Long-wavelength scalings}

At fixed radius, spatial Fourier transformation of Eq.~\eqref{eq:nu_memory_kernel_compact} gives
\begin{equation}
Z_m(\tau;R) = \frac1\chi \int\frac{\dd \bm k}{(2\pi)^d} |g'_R(k)|^2 e^{-\nu k^{2m}\tau}.
\label{eq:nu_memory_k_time}
\end{equation}
Similarly, Eq.~\eqref{eq:nu_noise_time} gives
\begin{equation}
B_{m,n}(\tau;R) = \frac{D}{\Gamma\chi} \int\frac{\dd \bm k}{(2\pi)^d} |g'_R(k)|^2 k^{2(n-m)} e^{-\nu k^{2m}|\tau|},
\label{eq:nu_noise_k_time}
\end{equation}
where $\tau=t-t'$.

The factor $g'_R(k)$ measures how strongly a displacement of the droplet wall couples to a bulk density mode of wave number $k$. In the thin-interface limit,
\begin{equation}
-g'_R(k)\simeq(2\pi)^{d/2}\Delta\rho R^{d-1} (kR)^{1-d/2}J_{d/2-1}(kR).
\label{eq:nu_shell_d}
\end{equation}
In particular,
\begin{equation}
-g'_R(0) =-\int\dd \bm r g'_R(r) \simeq\Om\Delta\rho R^{d-1},
\end{equation}
where $-g'_R(0)\delta R$ is the total mass added to the droplet by an infinitesimal increase in its radius.

At time $\tau$, modes satisfying $\nu k^{2m}\tau\gtrsim1$ have relaxed. It is therefore useful to introduce the propagation length
\begin{equation}
\ell(\tau)=(\nu \tau)^{1/(2m)}.
\end{equation}
For $\ell(\tau)\gg R$, the surviving modes have $kR\ll1$, so $g'_R(k)$ may be replaced by its value at $k=0$. Eqs.~\eqref{eq:nu_memory_k_time} and \eqref{eq:nu_noise_k_time} then give
\begin{align}
Z_m(\tau;R) &\propto R^{2d-2}\tau^{-d/(2m)},
\label{eq:nu_memory_tail}\\
B_{m,n}(\tau;R) &\propto R^{2d-2} \tau^{-[d + 2(n-m)]/(2m)}.
\end{align}

The memory is therefore integrable at long times for $d>2m$, marginal for $d=2m$, and nonintegrable for $d<2m$. Similarly, the projected noise correlation is integrable when
\begin{equation}
d>4m-2n.
\end{equation}
In particular, for $n=2m$ the noise correlation is integrable in every dimension considered here, even when the memory itself is long ranged. For $d=2$ and $m=1$, a memoryless theory predicts an unphysical logarithmic divergence of the response with system size~\cite{marqusee1984dynamics,zheng1989theory,cates2023classical}. An analogous divergence occurs in the two-dimensional Rayleigh--Plesset equation for an incompressible fluid. Accounting for compressibility removes the system-size divergence, replacing it with a long-time correlation~\cite{ilinskii2012models}.

For $d>2m$, the integrated memory defines a local drag coefficient,
\begin{equation}
\zeta_m(R) \equiv \int_0^\infty\dd\tau Z_m(\tau;R).
\end{equation}
Using Eq.~\eqref{eq:nu_shell_d}, the thin-shell result is
\begin{equation}
\begin{gathered} \zeta_m(R) =\frac{\Om(\Delta\rho)^2}{\Gamma} C_{d,m}R^{d + 2m-2},\\
C_{d,m}= \frac{\Gamma_E(2m-1)\Gamma_E(d/2-m)} {2^{2m-1}\Gamma_E(m)^2\Gamma_E(d/2 + m-1)}, \end{gathered}
\end{equation}
where $\Gamma_E$ denotes the Euler gamma function, distinct from the kinetic coefficient $\Gamma$.

When the noise correlation is integrable, we analogously define its long-time white-noise amplitude by
\begin{equation}
B_{m,n}(R) \equiv \frac{1}{2D} \int_{-\infty}^{\infty}\dd\tau B_{m,n}(\tau;R).
\label{eq:nu_noise_integrated}
\end{equation}
For the nonequilibrium branch $n=2m$,
\begin{equation}
B_{m,2m}(R)\simeq\frac{\Om R^{d-1}\gamma}{\kappa\Gamma^2}.
\end{equation}

For $n=2m-1$ and $d>2$,
\begin{equation}
B_{m,2m-1}(R) \simeq\frac{\Om(\Delta\rho)^2}{\Gamma^2(d-2)}R^d.
\end{equation}
In $d=2$, the correlation is marginal and decays as $|\tau|^{-1}$. Its effective strength therefore depends logarithmically on the observation time scale $T$. Defining
\begin{equation}
B_{m,2m-1}(T;R) \equiv \frac1D \int_0^\infty\dd\tau e^{-\tau/T} B_{m,2m-1}(\tau;R),
\end{equation}
one obtains, for $\ell(T)=(\nu T)^{1/(2m)}\gg R$,
\begin{equation}
B_{m,2m-1}(T;R) \simeq \frac{2\pi(\Delta\rho)^2R^2}{\Gamma^2} \left[ \ln\frac{\ell(T)}{R} + \ln2-\gamma_E \right],
\end{equation}
where $\gamma_E$ is the Euler--Mascheroni constant.

We next determine the response time of the critical droplet directly from the real-time equation. Write
\begin{equation}
R(t)=R_c + \delta R(t), \qquad \delta R(t)\propto e^{t/T_c}.
\end{equation}
To linear order, the radius dependence of the memory kernel may be frozen at $R_c$, since corrections to the kernel multiply ${\delta \dot R}$ and are therefore second order. Substituting the unstable mode into the radius equation gives
\begin{equation}
\frac1{T_c} \int_0^\infty\dd\tau Z_m(\tau;R_c)e^{-\tau/T_c} = |W''(R_c)| \propto R_c^{d-3}.
\label{eq:nu_unstable_mode}
\end{equation}
Using the long-time behavior in Eq.~\eqref{eq:nu_memory_tail} then gives
\begin{equation}
T_c\propto \begin{cases} R_c^{2m + 1},&d>2m,\\
R_c^{2m + 1}\ln(R_c/\ell_\gamma),&d=2m,\\
R_c^{2m(d + 1)/d},&2\leq d<2m, \end{cases} \quad \ell_\gamma=\frac{\chi\gamma}{(\Delta\rho)^2}.
\end{equation}

In two dimensions the weighted memory entering Eq.~\eqref{eq:nu_unstable_mode} can be evaluated explicitly. For $m=1$,
\begin{equation}
\int_0^\infty\dd\tau Z_1(\tau;R)e^{-\tau/T} \simeq \frac{2\pi(\Delta\rho)^2R^2}{\Gamma} I_0(z_T)K_0(z_T),
\end{equation}
with $ z_T={R}/{\sqrt{\nu T}}$. For $z_T\ll1$, $I_0(z_T)K_0(z_T)\simeq \ln(2/z_T)-\gamma_E$. At the critical droplet this gives
\begin{equation}
T_c=\frac{R_c^3\Lambda_c}{\nu \ell_\gamma}, \qquad \Lambda_c-\frac12\ln\Lambda_c =\frac12\ln\frac{R_c}{\ell_\gamma} + \ln2-\gamma_E,
\label{eq:nu_logarithm}
\end{equation}
or equivalently
\begin{equation}
\Lambda_c=-\frac12 W_{-1}\left( -\frac{e^{2\gamma_E}\ell_\gamma}{2R_c}\right),
\end{equation}
where $W_{-1}$ is the negative real branch of the Lambert function.

For $m>1$ in $d=2$,
\begin{equation}
\int_0^\infty\dd\tau Z_m(\tau;R)e^{-\tau/T} \simeq \frac{2\pi b_m(\Delta\rho)^2R^2} {\chi\nu ^{1/m}} T^{1-1/m},
\end{equation}
with $b_m={\pi}/{2m\sin(\pi/m)}$ and hence
\begin{equation}
T_c\simeq\frac{b_m^mR_c^{3m}}{\nu \ell_\gamma^m}.
\end{equation}

For the marginal even nonequilibrium branch $n=2m-1$ with $m\geq2$ in $d=2$, the noise logarithm on the critical-droplet time scale is
\begin{equation}
\mathcal L_c \equiv\ln\frac{\ell_c}{R_c} + \ln2-\gamma_E \simeq\frac12\ln\left(\frac{b_mR_c}{\ell_\gamma}\right)  + \ln2-\gamma_E,
\label{eq:nu_even_logarithm}
\end{equation}
with $\ell_c=(\nu T_c)^{1/(2m)}$.

\subsection{Activation barrier}

At equilibrium, the fluctuation--dissipation theorem implies
\begin{equation}
\Delta V_c^{\rm eq}=\Gamma\Delta W_c.
\end{equation}

Away from equilibrium the barrier depends on both the response and the projected noise. When $d>2m$, both are local on the slow droplet time scale. With $B_{m,n}(R)$ defined in Eq.~\eqref{eq:nu_noise_integrated}, the reduced one-dimensional action gives
\begin{equation}
\Delta V_c\simeq\int_0^{R_c}\dd R\frac{\zeta_m(R)W'(R)}{B_{m,n}(R)}.
\end{equation}
Thus explicit barrier expressions follow in this local limit. Using the thin-interface coefficients gives, for the odd branch $n=2m$,
\begin{equation}
\Delta V_{c} \simeq \frac{\Gamma\Om(d-1)\kappa(\Delta\rho)^2C_{d,m}} {(d + 2m-2)(d + 2m-1)}R_c^{d + 2m-2},
\end{equation}
and for the nonequilibrium even branch $n=2m-1$ with $m\geq2$,
\begin{equation}
\Delta V_{c} \simeq \frac{\Gamma\Om(d-1)(d-2)\gamma C_{d,m}} {(d + 2m-3)(d + 2m-2)}R_c^{d + 2m-3},
\end{equation}

At and below the memory threshold, the response is nonlocal. The marginal case $d=2$, $(m,n)=(1,2)$ admits the explicit leading-logarithmic estimate given below. More generally, the following scaling estimates assume that rescaling the entire escape history by $R_c$ and $T_c$ leaves a finite, nonzero minimized dimensionless action; the response-time calculation alone does not establish this assumption. For $n=2m$,
\begin{equation}
\Delta V_c\propto \begin{cases} R_c^{2d-2}\ln(R_c/\ell_\gamma),&d=2m,\\
R_c^{d-3 + 2m(d + 1)/d},&2\leq d<2m, \end{cases}
\end{equation}
whereas for $n=2m-1$ with $m\geq2$,
\begin{equation}
\Delta V_c\propto \begin{cases} R_c^{2d-3}\ln(R_c/\ell_\gamma),&d=2m,\\
R_c^{d-4 + 2m(d + 1)/d},&2<d<2m,\\
R_c^{3m-2}/\mathcal L_c,&d=2. \end{cases}
\end{equation}
For the case considered explicitly in the main text, $d=2$ and $(m,n)=(1,2)$, retaining the self-consistent logarithm rather than only its leading term $\tfrac12\ln(R_c/\ell_\gamma)$ gives
\begin{equation}
\Delta V_c \simeq\frac{\pi\Gamma\kappa(\Delta\rho)^2}{3} R_c^2\Lambda_c, \quad \Lambda_c=-\frac12 W_{-1}\left( -\frac{e^{2\gamma_E}\ell_\gamma}{2R_c}\right).
\label{eq:nu_m1_barrier}
\end{equation}

\section{Projection onto the slow interface mode}
\label{app:interface_constants}

We derive Eq.~\eqref{eq:interface_projected} directly from Eq.~\eqref{eq:model}, using the adjoint-mode projection of Ref.~\onlinecite{sarfati2026from} and the density decomposition defined in Eq.~\eqref{eq:interface_ansatz}. We Fourier transform along the interface, with tangential wave vector $\bm q$, and define $K_{\bm q}=|\bm q|^2-\partial_z^2$ and $H_g=f''(g)-\kappa\partial_z^2$. The planar (mean-field) profile obeys $f'(g)-\kappa g''=\mu_{\rm coex}$, so $H_g g'=0$. Linearization of Eq.~\eqref{eq:model} gives
\begin{equation}
\begin{gathered} \partial_t(-g'h_{\bm q} + \psi_{\bm q}) =\mathcal L_{\bm q}(-g'h_{\bm q} + \psi_{\bm q}) + \eta_{n,\bm q},\\
\mathcal L_{\bm q}=-\Gamma K_{\bm q}^m(H_g + \kappa |\bm q|^2). \end{gathered}
\label{eq:app_interface_linear}
\end{equation}
We follow Ref.~\onlinecite{sarfati2026from} and define $\ell_{\bm q}$ as the left eigenfunction of $\mathcal L_{\bm q}$: $\mathcal L_{\bm q}^\dagger\ell_{\bm q}=-\omega_{\bm q}\ell_{\bm q}$. We define $h(\bm x)$ by requiring $\langle\ell_{\bm q},\psi_{\bm q}\rangle=0$, where $\langle a,b\rangle=\int\dd z a^*b$. This condition assigns the selected capillary displacement to $h$ and the remaining density fluctuations to $\psi$. It removes the latter from the projected linear equation:
\begin{equation}
\langle\ell_{\bm q},\partial_t\psi_{\bm q}\rangle=0, \qquad \langle\ell_{\bm q},\mathcal L_{\bm q}\psi_{\bm q}\rangle =-\omega_{\bm q}\langle\ell_{\bm q},\psi_{\bm q}\rangle=0.
\end{equation}
In the long-wavelength limit, the leading adjoint mode is
\begin{equation}
\ell_{\bm q}\simeq K_{\bm q}^{-m}g',\qquad G_m(|\bm q|)=\langle g',K_{\bm q}^{-m}g'\rangle,
\end{equation}
since $H_gK_{\bm q}^m\ell_{\bm q}\simeq H_g g'=0$. This expression is asymptotic and the exact finite-$|\bm q|$ mode includes profile corrections. Unlike a change in the droplet radius, an interface deformation with a nonzero wave vector leaves the total phase volume unchanged at linear order, although it still requires local mass redistribution. The operator $K_{\bm q}=|\bm q|^2-\partial_z^2$ is then invertible, so the conserved zero-mode obstruction encountered in the nucleation projection is absent. Moreover, the slowest bulk relaxation rate at fixed $\bm q$ scales as $|\bm q|^{2m}$, while the capillary rate scales as $|\bm q|^{2m + 1}$. The bulk response therefore becomes quasistatic on the capillary time scale which justifies a local-in-time equation at leading order.

Taking the scalar product of Eq.~\eqref{eq:app_interface_linear} with this mode yields, to leading order,
\begin{equation}
\partial_t h_{\bm q}=-\frac{\Gamma\gamma |\bm q|^2}{G_m(|\bm q|)}h_{\bm q}  + \zeta_{\bm q},\qquad \zeta_{\bm q}=-\frac{\langle K_{\bm q}^{-m}g',\eta_{n,\bm q}\rangle}{G_m(|\bm q|)},
\end{equation}
where $\gamma=\kappa\int\dd z (g')^2$~\cite{onuki2002phase}. The bulk noise covariance $2D K_{\bm q}^n$ therefore gives
\begin{gather}
\begin{aligned} 
  \langle\zeta_{\bm q}(t)\zeta_{\bm q'}(t')\rangle
&=\frac{2D N_{m,n}(|\bm q|)}{G_m(|\bm q|)^2} (2\pi)^{d-1}\delta(\bm q + \bm q')\delta(t-t'), \end{aligned}\\
N_{m,n}(|\bm q|)=\int\frac{\dd p}{2\pi} |g'_p|^2(p^2 + |\bm q|^2)^{n-2m}.
\end{gather}
Here $p$ is the normal wave number. With $\Delta\rho=g(-\infty)-g( + \infty)$ and $C_a=\Gamma_E(a-\tfrac12)/(2\sqrt\pi\Gamma_E(a))$, rescaling $p=|\bm q| u$ gives
\begin{align}
G_m(|\bm q|)&\simeq(\Delta\rho)^2C_m |\bm q|^{1-2m},\\
N_{m,n}(|\bm q|)&\simeq \begin{cases} (\Delta\rho)^2C_{2m-n}|\bm q|^{2n-4m + 1},&n\leq2m-1,\\
J_{m,n},&n\geq2m, \end{cases}
\end{align}
where $J_{m,n}=\int\dd p |g'_p|^2p^{2n-4m}/(2\pi)$ is finite for a smooth profile. Thus
\begin{gather}
\nu ^{h}=\frac{\Gamma\gamma}{(\Delta\rho)^2C_m},\\
\mathcal D_{m,n}^{h}=\begin{cases} \dfrac{D C_{2m-n}}{(\Delta\rho)^2C_m^2},&n\leq2m-1,\\[3pt]
\dfrac{D J_{m,n}}{(\Delta\rho)^4C_m^2},&n\geq2m, \end{cases}\\
\alpha_{m,n}=\begin{cases} 2n-1,&n\leq2m-1,\\
4m-2,&n\geq2m, \end{cases}
\end{gather}
which gives Eq.~\eqref{eq:interface_projected}.

\subsection{How to measure \texorpdfstring{$h$}{h}}
\label{app:howto}
To compare height definitions, we write $\delta\rho=\rho-g$ and consider two different measurements of $h(\bm x)$:
\begin{equation}
h_{\mathrm{w},\bm q} =-\frac{\langle K_{\bm q}^{-m}g', \delta\rho_{\bm q}\rangle} {\langle g',K_{\bm q}^{-m}g'\rangle}, \qquad h_{\mathrm{fit},\bm q} =-\frac{\langle g',\delta\rho_{\bm q}\rangle} {\langle g',g'\rangle}.
\end{equation}
The first is the leading-order form of the projected definition introduced in the previous subsection. The second is the least-squares estimate of a small translation of the planar profile, with its width and amplitude fixed (a fit to the mean profile at each slice essentially). 

When their spectra diverge at small $q$, the two height measurements typically agree. However, when the weighted definition gives vanishing fluctuations at small $q$, the local profile-fit definition can give a finite, nonzero spectral background, because density fluctuations near the interface also shift the fitted position. This is what we observe over the measured range in Fig.~\ref{fig:interface}(c).

\bibliography{main}

\end{document}